Molecular Dynamics Studies on the Structural Stability of Wild-Type Rabbit Prion Protein: Surface Electrostatic Charge Distributions

Jiapu Zhang
*School of Sciences, Information Technology and Engineering, University of Ballarat, Mount Helen, Ballarat, VIC 3350, Australia, Emails: j.zhang@ballarat.edu.au , jiapu_zhang@hotmail.com
Phone: (61)423 487 360.*

**Abstract**
Prion diseases cover a large range of neurodegenerative diseases in humans and animals, which are invariably fatal and highly infectious. By now there have not been some effective therapeutic approaches or medications to treat all prion diseases. Fortunately, numerous experimental experiences have showed that rabbits are resistant to infection from prion diseases isolated from other species, and recently the molecular structures of rabbit prion protein and its mutants were released into protein data bank. Prion diseases are "protein structural conformational" diseases. Thus, in order to reveal some secrets of prion diseases, it is amenable to study rabbits by techniques of the molecular structure and its dynamics. Wen et al. (PLoS One 5(10) e13273 (2010), Journal of Biological Chemistry 285(41) 31682-31693 (2010)) reported the surface of NMR RaPrP$^C$(124-228) molecular snapshot has a large land of continuous positive charge distribution, which contributes to the structural stability of rabbit prion protein. This is just the property at one snapshot. This paper confirms a large land of positive charge distribution has still been reserved during the long molecular dynamics of 30 nanoseconds in many environments, but the continuous of the land has not been always reserved yet. These results may be useful for the medicinal treatment of prion diseases.

# 1 Introduction

Prion diseases are invariably fatal and highly infectious neurodegenerative diseases affecting humans and animals. The neurodegenerative diseases such as Creutzfeldt-Jakob disease (CJD), variant Creutzfeldt-Jakob diseases (vCJD), iatrogenic CreutzfeldtJakob disease (iCJD), familial CreutzfeldtJakob disease (fCJD), sporadic CreutzfeldtJakob disease (sCJD), Gerstmann-Straussler-Scheinker syndrome (GSS), Fatal Familial Insomnia (FFI), KURU in humans, Scrapie in sheep, bovine spongiform encephalopathy (BSE or mad-cow disease) in cattle, chronic wasting disease (CWD) in white-tailed deer, elk, mule deer, moose, transmissible mink encephalopathy (TME) in mink, feline spongiform encephalopathy (FSE) in cat, exotic ungulate encephalopathy (EUE) in nyala, oryx, greater kudu, and spongiform encephalopathy (SE) in ostrich etc belong to prion diseases (http://en.wikipedia.org/wiki/Prion). By now there have not been some effective therapeutic approaches or medications to treat all these prion diseases.

Rabbits are one of the few mammalian species reported to be resistant to infection from prion diseases isolated from other species [1, 2, 3, 4, 5, 6, 7, 8, 9]. Thus, the paper studies rabbit prion protein in order to get some clues for the treatment of prion diseases.

Prion diseases are "protein structural conformational" diseases. The normal cellular prion protein (PrP$_C$) is rich in α-helices but the infectious prions (PrP$_{Sc}$) are rich in β-sheets amyloid fibrils. The conversion of PrP$_C$ to PrP$_{Sc}$ is believed to involve a conformational change from a predominantly α-helical protein (about 42% α-helix and 3% β-sheet) to a protein rich in β-sheets (about 30% α-helix and 43% β-sheet) [10]. Fortunately, the X-ray and NMR protein structures of rabbit prion protein and its S173N and I214V mutants were released into protein data bank (www.rcsb.org) (PDB ID codes 3O79, 2FJ3, 2JOH, 2JOM) recently. Hence, the conformational changes may be amenable to study by molecular dynamics (MD) techniques of these X-ray and NMR structures.

The author has studied the structural dynamics of rabbit prion protein and found that the structures were unfolded from α-helices into β-sheets under the conditions of 350 K or 450 K in low pH environment [11, 12, 13, 14, 15, 16]. However, under these conditions, the α-helices of human, mouse, dog and horse prion proteins were not found to unfold [11, 12, 13, 14, 15, 16]. The author concluded that some salt bridges (e.g. the one between residues 163 and 177) are clearly contributing to the structural conformational change [11, 12, 13, 14, 15, 16]. β2-α2 loop was reported to play an important role to stabilize the structural stability of rabbit and horse prion proteins [17, 18, 19, 20, 21, 22, 23, 24, 25, 26, 27, 28], and recently Wen et al. (2010) reported that rabbit prion protein has a unique distribution of large-continuous-positive -charge-surface electrostatic potential [25, 24, 21], which is with the highly ordered and well NMR-signal-recognized β2-α2 loop contributing to the structural stability of rabbit prion protein. The paper will study (by MD techniques) the dynamics of the surface electrostatic potential distributions of the 3D NMR structure of rabbit prion protein.

## 2 Surface Electrostatic Charge Distributions

To describe the potential energy associated with a charge distribution the concept of the electrostatic potential is introduced. The electrostatic potential at a given position is defined as the potential energy of a test particle divided by the charge of this object. When the charge distribution is over a particular area then the distribution is called as surface charge distribution. Electrostatic potential surfaces are valuable in computer-aided drug design because they help in optimization of electrostatic interactions between the protein and the ligand. These surfaces can be used to compare different inhibitors with substrates or transition states of the reaction.

First, we show the surface electrostatic charge distributions of 3D NMR structures of rabbit, horse, dog, human, mouse, rabbit S173N mutant and rabbit I214V mutant prion proteins (Figure 1).

< Figure 1 >

We may see in Figure 1 that wild-type rabbits have a larger land of positive charge distributions than horses, dogs, humans and mice. Figure 1 just illuminates to us the surface charge distribution of one snapshot of rabbit prion protein. In this paper we will study the dynamics of the surface charge distributions during long MD simulations of 30 ns to confirm the above observation.

## 2.1 MD Techniques

MD simulation materials are completely same as the ones of [11, 12, 13, 14] and the initial simulation structures were built on RaPrP$_C$(124-228) (PDB entry 2FJ3). MD simulation methods are completely same as the ones of [11, 12, 13, 14]. The next subsection illuminates the surface charge distributions of RaPrP$_C$ on the snapshots at 5 ns, 10 ns, 15 ns, 20 ns, 25 ns, 30 ns of 350 K and 450 K for different initial velocities seed 1 and seed 2, in neutral and low pH environments (Figures 2-9).

## 2.2 Surface Charge Distributions of RaPrP

< Figure 2 >

< Figure 3 >

< Figure 4 >

< Figure 5 >

< Figure 6 >

< Figure 7 >

< Figure 8 >

< Figure 9 >

## 2.3 Analyses of Surface Charge Distributions of RaPrP

We may see in Figures 2-9 that the continuation of blue or red colors for low pH value is less for neutral pH value; this might imply to us RaPrP has less structural stability in low pH environment. Similarly, we may list the snapshots of surface charge distributions of HoPrP$^C$(119–231) (PDB entry 2KU4), DoPrP$^C$(121–231) (PDB entry 1XYK), HuPrP$^C$(125-228) (PDB entry 1QLX) and MoPrP$^C$(124-226) (PDB entry 1AG2) during the long MD simulations of 30 ns. However, we cannot find their clear differences from RaPrP$^C$(124-228). In the below, we only analyze the surface charge distributions of special residues ARG163 (Figure 10) and ASP177 (Figure 11) and of the β2-α2 loop RaPrP(163-171) (Figure 12).

< Figure 10 >

< Figure 11 >

< Figure 12 >

By observations from Figure 12 we cannot see much difference among these snapshots at the 30th nanosecond. Figures 10-11 illuminate to us that the NMR snapshot has large land of uniform distributions around the the residues R163 and D177 respectively, but we cannot find the real difference among all the snapshots at the 30$^{th}$ nanosecond.

## 3 Conclusion

The surface of NMR RaPrP$_C$(124-228) molecule has a large land of continuous positive charge distribution (Figure 13). This agrees with the observation of [25, 24, 21]. To check whether this continuous positive charge distribution is still reserved during the long MD simulations of 30 ns under many different conditions, this paper presents numerous snapshots of graphs for NMR RaPrP(124-228). These graphs should have a value acted as a rich database to study the 'unique distribution of surface electrostatic potential' of NMR RaPrP$_C$(124-228) molecule.

< Figure 13 >

Generally, we find that, in neutral pH environment, rabbit prion protein has a large land of positive charge distribution (Figures 1a, 1g, 3, 5, 7, 9, 10a, 11a); this agrees with the finding of Wen et al. [25, 24, 21]. In low pH environment, the large continuous and disappears (Figures 2, 4, 6, 8) and the structural stability of rabbit prion protein is broken (from α-helices into β-sheets). In conclusion, the surface electrostatic potential charge distribution, more or less, contributes to the structural stability of rabbit prion protein and contributes to the study of the species barrier of rabbit prion protein to prion diseases.

## Acknowledgments
This research was supported by a Victorian Life Sciences Computation Initiative (VLSCI) grant number VR0063 on its Peak Computing Facility at the University of Melbourne, an initiative of the Victorian Government.

**Figures 1-13** can be seen from the last attached PDF file in website

http://blog.51xuewen.com/jiapu_zhang_PhD_MSc_MSc_BSc/article_43241.htm